\documentclass[%
twocolumn,
 amsmath,amssymb,
 aps,
 showpacs
]{revtex4-1}
\usepackage{amsmath}
\usepackage{mathrsfs}
\usepackage{graphicx}
\usepackage{float}
\usepackage{dcolumn}
\usepackage{bm}
\usepackage{ulem}
\usepackage[
margin=.75in,
]{geometry}

\begin{document}
\title{Generation of isolated attosecond pulses by polarization gating of high-order harmonic emission from H$_2^+$ in intense ultrashort laser fields}

\author{Nehzat Safaei}

\email {Corresponding author: safaei@ut.ac.ir}

\affiliation{
Department of Physical Chemistry, School of Chemistry, College of Science, University of Tehran, Tehran, I. R. Iran
 }

\begin{abstract}
Quantum-mechanical investigation is presented to study single attosecond pulse generation by gating high-order harmonic emission from H$_2^+$ molecule in intense laser pulses with time-dependent ellipticity. The high-order harmonic generation from H$_2^+$ molecule in superposition of a left and a right-hand circularly polarized Gaussian pulse is studied and the effect of time duration and carrier-envelope phase of laser field on single attosecond pulse generation is investigated. Using laser field formed by  combination of a left and a right-hand elliptically polarized Gaussian pulse, the effect of ellipticity of field on duration and intensity of generated attosecond pulses is studied. The numerical calculations show that, with the conventional polarization gating, an intense single attosecond pulse can be isolated from the pulse train emitted by H$_2^+$ molecule in the laser field.
\end{abstract}

\maketitle
\section{Introduction}
 High-order harmonic generation (HHG) from atoms and molecules in intense laser fields has been proved to be an important technique for producing coherent high-energy attosecond pulse trains as well as single isolated attosecond pulses [1-3]. The mechanism of HHG process can be  interpreted by a three-step semiclassical model [4-6]: first, under an intense laser field, the bound state electron tunnels through the coulomb barrier into continuum, then the freed electron  accelerates in the external laser field and finally it may recombine into the ground state due to the phase change of the electric field followed by attosecond burst of electromagnetic waves emission. In many cycle laser fields this process occurs every half-cycle and generates attosecond pulse trains which due to the interference of these attosecond pulses there is a frequency comb with an interval of twice the driving field photon energy.
In recent years, generation of isolated pulses with very short duration has gained enormous interest and much effort has been paid to achieve single attosecond pulses with long pulse lasers as they are easier to access. For example, it has been proposed to manipulate two-color laser fields [7-11] so that a continuum spectrum in the cutoff region of HHG spectra can be generated. Another alternative approach to generate isolated attosecond pulses is using of polarization gating of ultrashort pulses as suggested by Corkum and his collaborators [12].
The polarization gating is based on the dependence of the high harmonic generation efficiency on the ellipticity of the laser. The gating is applies with laser pulses whose ellipticity changes from circular to linear and conversely. In this created field, the electron will be driven away from the nucleus, in the head and tail parts, and in the center portion of the pulse, which is defined as the polarization gate, single attosecond pulses can be generated without spatial filtering. 
Chang [13] has investigated polarization gating in atoms via semiclassical theory. This method  mainly takes ground and continuum states into account and both the ionization and recombination processes occur in the ground state. 
However, recent studies have shown that not only the ground state but the excited states contribute to the HHG spectrum [14-15]. 
On the other hand, HHG from molecules shows some differences compared to that from atomic systems, such as two-center interference and a double-plateau structure [16-17]. Typical molecular HHG spectra exhibit a narrow and a broad plateau at low and high frequencies respectively. In the broad part of the plateau, the efficiency  exceeds that of the atomic-like one by several orders of magnitude. 
In molecular high-order harmonic generation, the ionized electron can recombine with neighboring ions in the same molecule leading to a much larger cutoff energy depending on the internuclear distance. This larger cutoff energy comes from longer acceleration distance in the recombination process. With the advantage of the high intensity of the molecular plateau and larger cutoff energy, the  molecular HHG could be an important topic of research to achieve an isolated attosecond laser pulse. Ge \textit{at al.} [18] have shown that an isolated attosecond pulse with duration 129 as could be generated from the H$_2^+$ molecule in 3-fs, 800-nm laser pulses.
Also, recent studies proposed that, single circularly and elliptically attosecond pulses, which are potential new tools for investigation of electron dynamics inside atoms and molecules, could be produced by molecules exposed to the intense laser pulses with a time-dependent ellipticity [19-20]. Yuan and Bandrauk [21] have been generated a single circularly polarized 114 as pulse from the interaction of H$_2^+$ molecule with an elliptically polarized laser pulse at a wavelength of 400 nm in the presence of a Terahertz Field.\\ 
 \hspace*{3 mm} In this paper, the quantum approach is performed to investigate polarization gating of HHG from H$_2^+$ molecule by solving the time-dependent Schr\"{o}dinger equation (TDSE) numerically, which provides us exact results. 
\section{Theory and Computational methods}
The interaction of the  H$_2^+$ molecule with a laser field with time-dependent degree of ellipticity is described by the corresponding two-dimensional time-dependent Schr\"{o}dinger equation as [22-24]: (atomic units (a.u.) are used throughout the paper unless stated otherwise.)
\begin{eqnarray}\label{eq:11}
  i \frac{\partial \psi(x,y,t)}{\partial t}=[{\widehat {H}_0(x,y)}+\widehat H(x,y,t) ]\,\psi(x,y,t),
\end{eqnarray}
where the unperturbed Hamiltonian $\widehat{H}_0(x,y)$ of the system is given by:                                   
\begin{eqnarray}\label{eq:12}
 \widehat{H}_0(x,y)=-\frac{1}{2}\nabla_{x,y}^2+ \widehat{V}(x,y).
\end{eqnarray}
In Eq. (2), $\widehat{V}(x,y)$ is the soft Coulomb potential of the system:
\begin{eqnarray}\label{eq:3}
 \widehat{V}(x,y)=\sum_{i=1}^{2} -1/{\sqrt{\beta+(x-x_i)^2+(y-y_i)^2}}.
\end{eqnarray}
In the above equation ($x_i$,$y_i$) are coordinates of nucleus in the molecule and $\beta$ is the softening parameter which is chosen 0.52 to produce the real energy curve of the 1$\sigma_g$ state of the H$_2^+$ molecule. ͯDuring the simulations an absorbing potential included to avoid unphysical reflections of the electron wavepacket at the boundaries.
The interaction term in length gauge is $H(x,y,t) = E(t).(x \, \widehat{x} +y\, \widehat{y})$. 
In the present paper, $E(t)= E_x(t)\, \widehat{x} + E_y(t)\, \widehat{y}$, is considered as a polarized pulse with a time-dependent ellipticity, which is generated by the superposition of a left and a right-hand polarized Gaussian pulse. The electric fields of the left and right-hand polarized pulses propagating in the z direction are:
\begin{eqnarray}\label{eq:3}
 E_l(t)= E_0\, e^{-2 Ln(2)({(t-t_d/2)/\tau_p)}^2}[\widehat{x}\,cos(\omega t +\phi)\\
 \nonumber
 +\, \widehat{y}\,\epsilon\,sin(\omega t +\phi)](-1)^n,\\
 E_r(t)= E_0\, e^{-2 Ln(2)({(t+t_d/2)/\tau_p)}^2}[\widehat{x}\,cos(\omega t +\phi)\\
 \nonumber
 -\, \widehat{y}\,\epsilon\,sin(\omega t +\phi)](-1)^n.
\end{eqnarray}
 The peak field amplitude $E_0$, carrier frequency $\omega $, pulse duration $\tau_p $ and carrier-envelope phase $\phi $ are the same for the two pulses and $t_d$, which is an integral number, $n$, of optical periods, is the time delay between left and right-hand pulses.
 In case of $ \epsilon=1$, the electric field components are left and right-circularly polarized pulses propagating in the z direction. If $ \epsilon\neq 1$ the field is a superposition of a left and a right- elliptically polarized fields.
The TDSE is solved using unitary split-operator technique where an eleven-point finite difference method is used for calculating the first and second derivatives. The Crank-Nicolson method which expresses the exponential operator to the third order is used to handle the time propagation.
 Simulation boxes are set to 200 a.u. $\times$ 200 a.u. and the adaptive grid spacing is set to 0.2 a.u. (near the center of the simulation box) and 0.5 a.u. (near the borders of the simulation box) in both directions. The corresponding time step is set to be 0.01 a.u. and the internuclear distance is fixed at R = 2 a.u..
Based on the Ehrenfest theorem, the time-dependent dipole acceleration in each direction can be read as [25]:
 \begin{eqnarray}\label{eq:3}
 a_x(t)=\langle \psi(x,y,t)\vert\,\widehat{x}.[\nabla V(x,y)+E(t)]\,\vert\psi(x,y,t)\rangle \nonumber \\
 a_y(t)=\langle \psi(x,y,t)\vert\,\widehat{y}.[\nabla V(x,y)+E(t)]\,\vert\psi(x,y,t)\rangle
\end{eqnarray}
The x and y components of profile of the attosecond pulses can be obtained  via superposing several harmonics
 \begin{flalign}
I_x(t) =\vert \sum_q a_q \, exp[iq\omega t] \vert ^2,\nonumber \\
I_y(t) = \vert \sum_q a^\prime _q \, exp[iq\omega t] \vert ^2,
 \end{flalign}
 where
  \begin{flalign} 
 a_q=\int a_x(t)exp[-iq\omega t] dt ,\nonumber \\
  a^\prime _q=\int a_y(t)exp[-iq\omega t] dt.
  \end{flalign}
The time dependence of harmonics is obtained by Morlet wavelet transform of dipole acceleration as:
\begin{eqnarray}\label{eq:7}
  &w(\omega,t)= \sqrt{ \frac{\omega}{\pi^\frac{1}{2}\sigma}}\times
 \nonumber \\
 \vert &\int_{-\infty}^{+\infty}a_x(t^\prime)\,\widehat{x}\,exp[-i\omega (t^\prime-t)]exp[-\frac{\omega^2 (t^\prime-t)^2}{2\sigma^2}]dt^\prime
 +  \nonumber \\ &\int_{-\infty}^{+\infty}a_y(t^\prime)\,\widehat{y}\,exp[-i\omega (t^\prime-t)]exp[-\frac{\omega^2 (t^\prime-t)^2}{2\sigma^2}]dt^\prime  \vert.
\end{eqnarray}
\begin{figure*}[ht]
\centering
\resizebox{120mm}{100mm}{\includegraphics[bb=13 0 429 407]{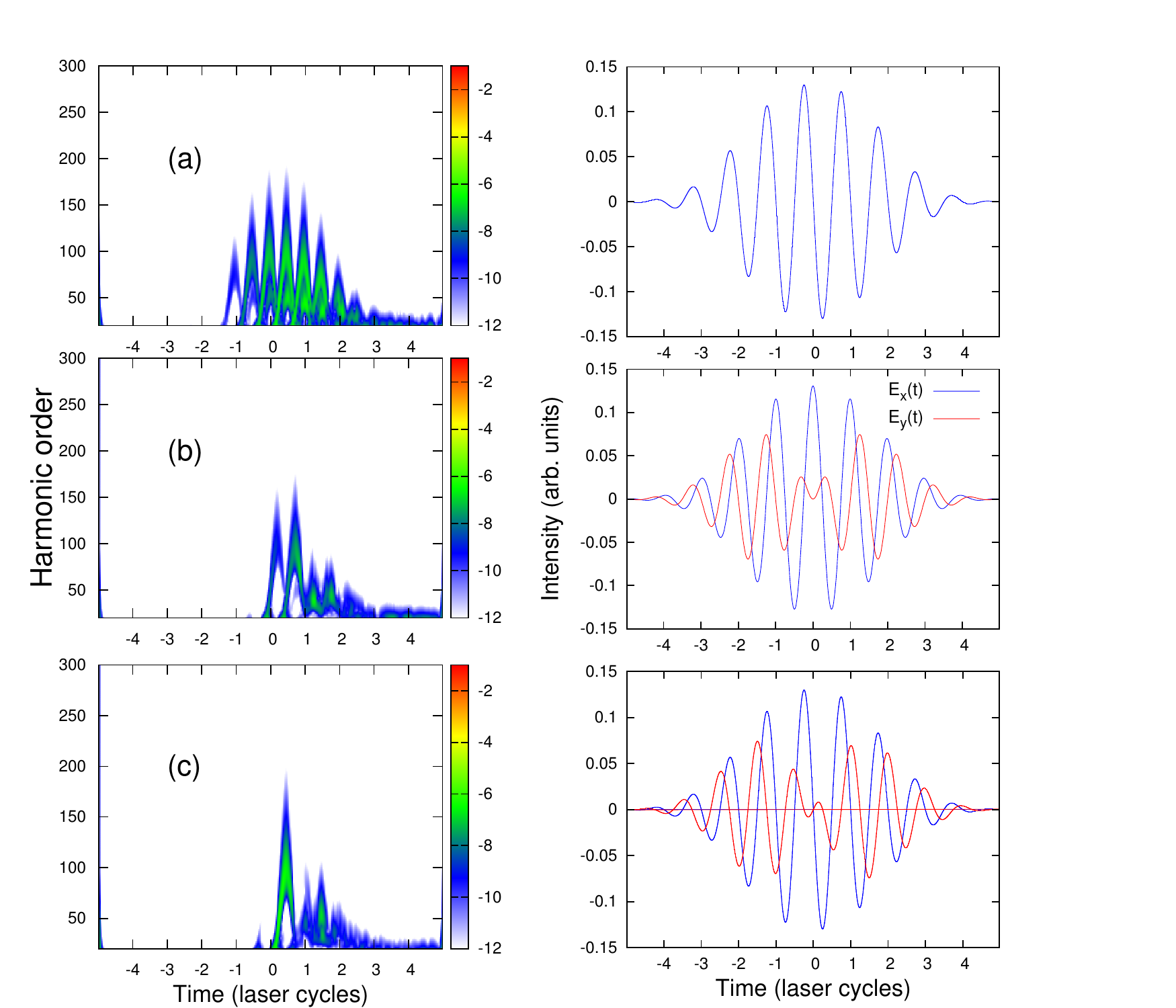}}
\caption{
(Color online) The Morlet wavlet time profiles (left panels) for H$_2^+$ molecule in different laser field shapes (a)-(c)(right panels), formed by combination of a left and a right-hand circularly polarized Gaussian pulse with $\tau_p=T_d=220$ a.u. ($\sim$ 5 fs) at 800 nm wavelength ($\omega=0.057$ a.u.) and $I$=3$\times 10^{14}$ W/cm$^2$ intensity: (a) x component of the laser field with carrier-envelope phase $\phi =0$ (b) laser field with carrier-envelope phase $\phi =0$ (c) laser field with carrier-envelope phase $\phi =\pi/2$.\label {fig:q}}
\end{figure*}
\begin{figure}[]
\begin{tabular}{l}
\resizebox{60mm}{65mm}{\includegraphics[bb=5.922000 6.966000 205.973994 312.200006]{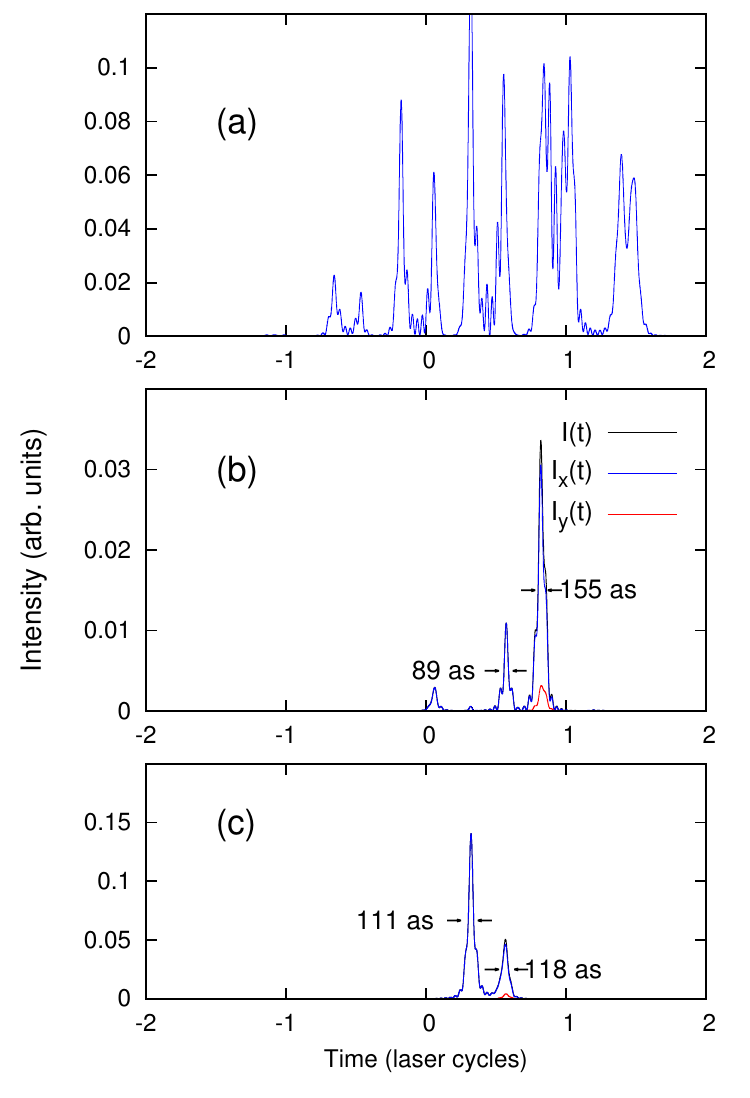}}
\end{tabular}
\caption{
(Color online) The profiles of the generated attosecond pulses from the high-order harmonic spectrum of H$_2^+$ molecule in different laser field shapes (a)-(c), formed by combination of a left and a right-hand circularly polarized Gaussian pulses with $\tau_p=T_d=220$ a.u. ($\sim$ 5 fs) at 800 nm wavelength ($\omega=0.057$ a.u.) and $I$=3$\times 10^{14}$ W/cm$^2$ intensity: (a) the x component of laser field with carrier-envelope phase $\phi =0 $ (b) laser field with carrier-envelope phase $\phi =0 $ (c) laser field with carrier-envelope phase $\phi =\pi/2$.\label {fig:q}}
\end{figure}
\begin{figure*}[ht]
\centering
\resizebox{100mm}{90mm}{\includegraphics[bb=4.122000 3.582000 419.949972 347.003989]{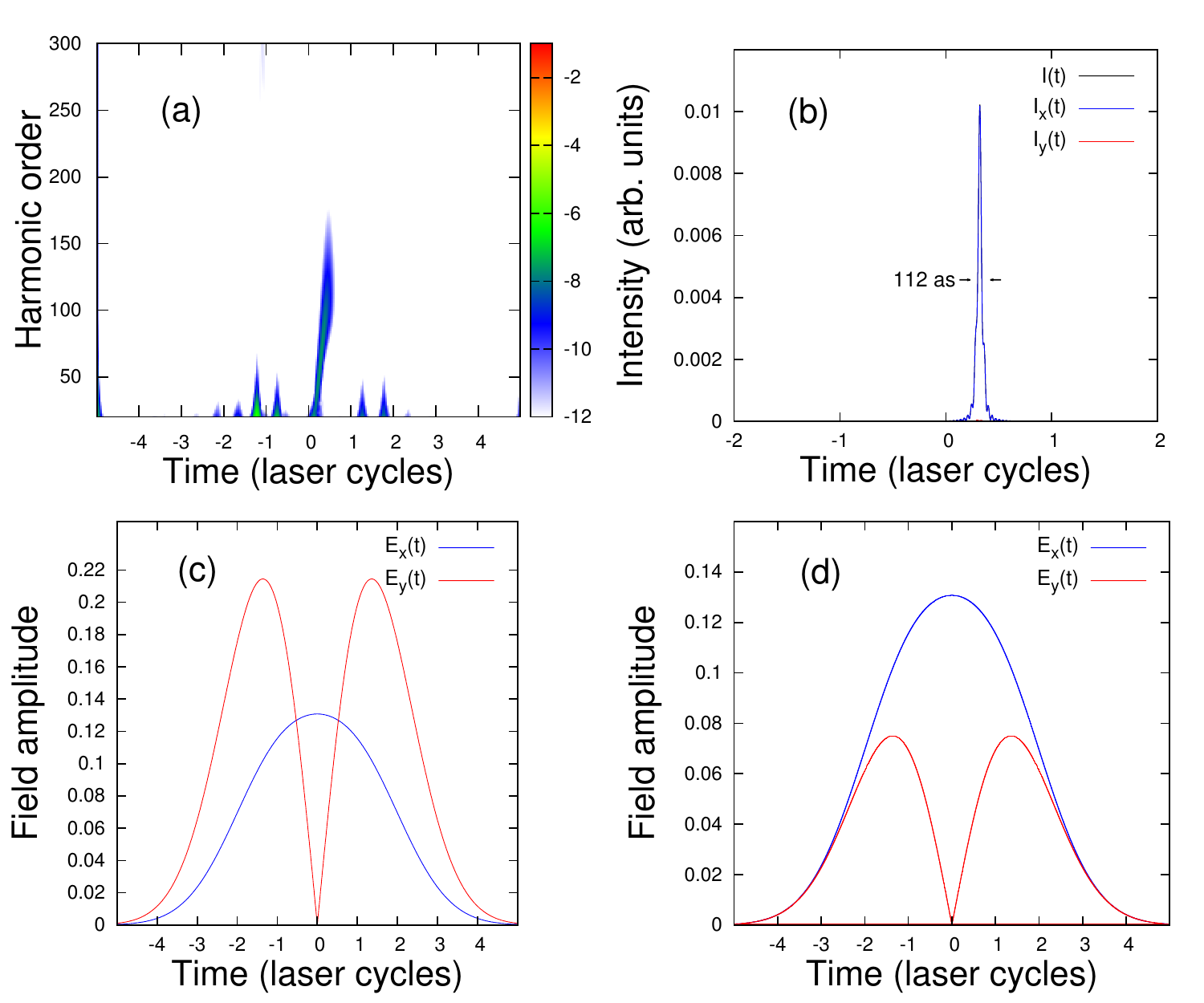}}
\caption{
(Color online)(a) The Morlet wavlet time profiles for H$_2^+$ molecule in superposition of a left and a right-elliptically polarized Gaussian laser pulse with $\tau_p=T_d=220$ a.u. ($\sim$ 5 fs) and $\epsilon=0.35$ at 800 nm wavelength ($\omega=0.057$ a.u.) and $I$=8.6 $\times 10^{14}$ W/cm$^2$ intensity. (b) The profile of the extracted attosecond pulse from the HHG spectrum. (c) Field amplitude of x and y components of elliptically polarized laser pulses ($\epsilon=0.35 $). (d)  Field amplitude of x and y components of circularly polarized laser pulses ($\epsilon=1 $) with $I$=3$\times 10^{14}$ W/cm$^2$ intensity.\label {fig:q}}

\end{figure*}
\vspace*{-7 mm}
\section{Results and Discussion}
To investigate the polarization gating for generation of isolated attosecond pulses, first the Schr\"{o}dinger equation is solved numerically for interaction of H$_2^+$ molecule with a laser field created by superposition of a left and a right-hand circularly polarized Gaussian pulses, using the method described in Sec II. The time-dependent ellipticity of input pulse takes the form of [26]:
\begin{eqnarray}\label{eq:3}
 \xi(t)=-\frac{\vert 1-e^{-4 Ln(2)((t_d/{\tau_p}^2)t}\vert}{1+e^{-4 Ln(2)((t_d/{\tau_p}^2)t}}.
\end{eqnarray}
 For the center portion of the pulse where the field is almost linearly polarized i.e., $-4 Ln(2)((t_d/{\tau_p}^2)t<<1$, $\xi(t)$ can be expressed as:
 \begin{eqnarray}\label{eq:3}
 \xi(t)\approx\vert{-4 Ln(2)((t_d/{\tau_p}^2)t}\vert.
\end{eqnarray}
 From the above equation, the time interval, where the ellipticity is less than a certain value $\xi(t)$, is thus:
\begin{eqnarray}\label{eq:3}
 \delta(t)=\frac{1}{Ln(2)}\xi(t)\frac{{\tau_p}^2}{T_d}
\end{eqnarray}
Using Eq. (12) the width of the gate part where the pulse contributes to the harmonic generation  effectively i.e., $\xi<0.2$, can be calculated as: 
 \begin{eqnarray}\label{eq:3}
 \delta(t)=0.3 \frac{{\tau_p}^2}{T_d}.
\end{eqnarray}
According to  Eq. (12) there are two ways to reduce the high harmonic radiation time and to make the gate shorter than one optical cycle. The gate width reduces by using either shorter input pulses or a longer delay time. Since the radiation time depends on the square of the pulse duration, using shorter pulses is more effective but it is limited by the shortest laser pulse that could be generated. In the latter approach, most of the laser energy is outside the gate so the conversion efficiency is low. 
In the present paper, the polarization gating technique is applied to control the HHG process and to generate isolated attosecond laser pulses for a compromised condition where $\tau_p=T_d$.\\
\hspace*{3 mm}First the calculations are performed for the superposition of a left and a right-hand circularly polarized pulse with 220 a.u. ($\sim$ 5 fs) time duration and separated by 220 a.u.. 
Laser pulses centered at 800 nm, and the peak intensity set to $I$=3$\times 10^{14}$ W$/$cm$^2$. The calculated time profiles of high harmonic emission and corresponding field components are presented in Figs. 1(a)-1(c). For the sake of comparison with harmonic generation from interaction of H$_2^+$ molecule with linearly polarized laser pulses, the time profile is calculated for the case where only the driving part of the pulse is taken into account and the result is presented in Fig. 1(a). Figs. 1(b) and 1(c) show the results for HHG in perpendicularly polarized pulses with carrier-envelope phase of 0 and $\pi/2 $ respectively.\\
According to the right panels of Fig. 1(b) and 1(c), the x component of field ($E_x(t)$) is driving field and it is responsible for the generation of attosecond pulses whereas the y component of field ($E_y(t)$) is the gating part, which suppresses the harmonic emission in the gate.
As it can be seen from the Fig. 1(a), a train of pulses, separated by half a laser cycle, is obtained from the interaction of H$_2^+$ molecule with the linearly polarized field. For polarization gating with laser pulse with $\phi=0$, Fig. 1(b), four pulses are left while only two pulses with comparable intensity are survived in the case of $\phi=\pi/2$. There are pulses, at low frequencies, outside the gate in both case of $\phi=0$ and $\phi=\pi/2$, which it is assumed that after reaching the maximum of laser pick amplitude, where quantum state of the system are changed extremely, due to the quantum diffusion of electron wavepacket the gate width is larger than the calculated width from the Eq. (12). 
The calculated attosecond pulses for initial laser field with different gating schemes, created by superposing the harmonics of plateau from the 60th to the 90th orders, are shown in Fig. 2(a)-2(c). It can be seen from Fig. 2(a), when  H$_2^+$ molecule was pumped by linearly polarized field, a train of ten attosecond pulses is generated, while the number of pulses is reduced to two in laser pulses with $\phi=\pi/2$ with the conventional polarization gating, which being farther from the gate, one of them has a lower intensity. In the case of $\phi=0$ three pulses are left. 
The separation between the generated pulses is the same as for linearly polarized pump pulses.
The results, in agreement with Eq. (12), show that when the H$_2^+$ molecule is pumped by superposition of circularly polarized pulses, separated by a delay, the HHG is gated.\\
\hspace*{3 mm}In this part, the interaction of H$_2^+$ molecule with superposition of a left and a right-elliptically polarized Gaussian laser pulses is studied. In this case, the gate width is:
 \begin{eqnarray}\label{eq3}
 \delta(t)=0.3\, \epsilon\, \frac{{\tau_p}^2}{T_d}.
\end{eqnarray}
\begin{figure*}[ht]
\resizebox{100mm}{120mm}{\includegraphics[bb=10 24 440 550]{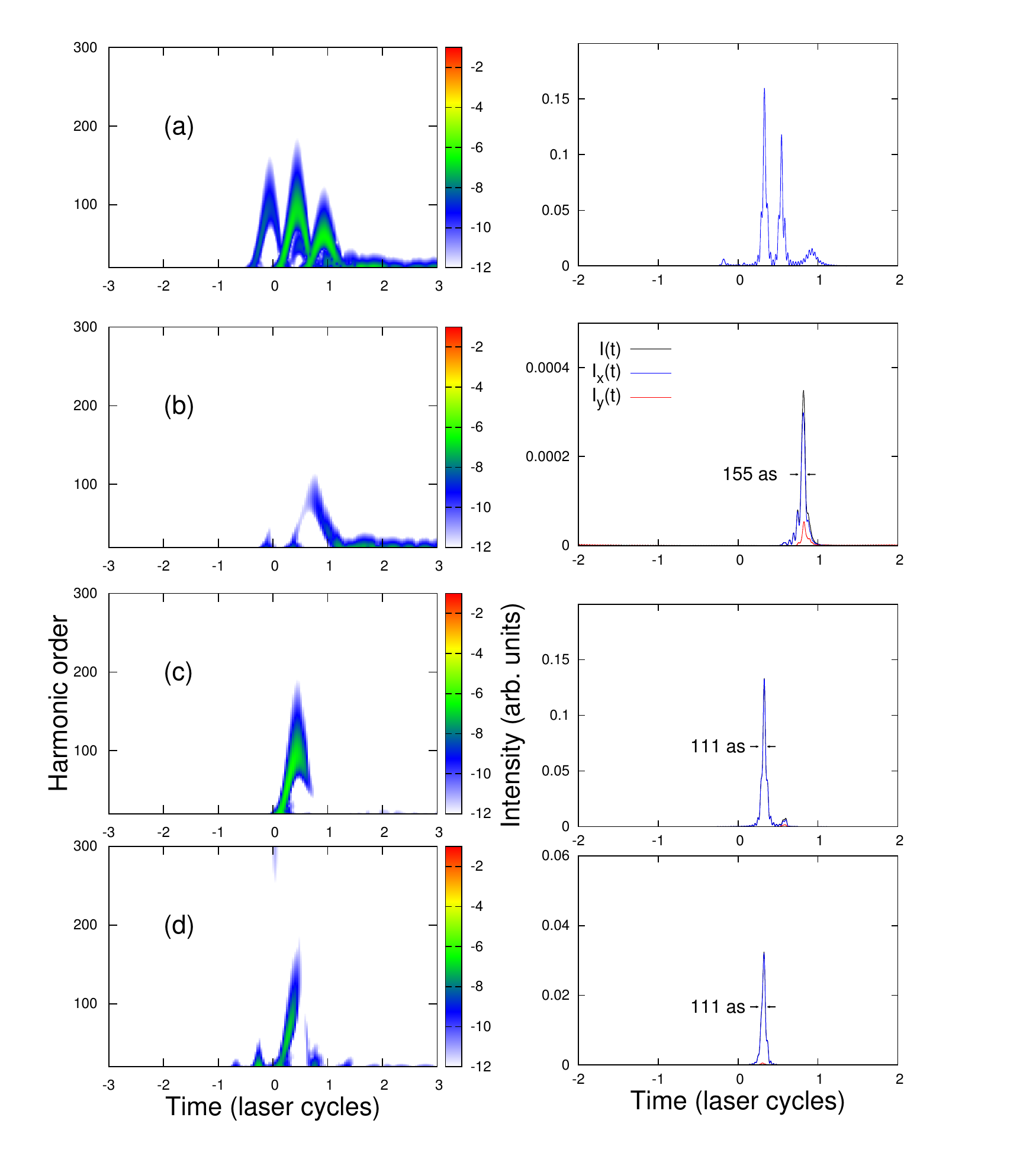}}
\caption{
(Color online) The Morlet wavlet time profiles (left panels) for H$_2^+$ molecule in different laser field shapes (a)-(c), formed by combination of a left and a right-hand circularly polarized Gaussian pulse with $\tau_p=T_d=110$ a.u. ($\sim$ 2.5 fs) at 800 nm wavelength ($\omega=0.057$ a.u.), $I$=3$\times 10^{14}$ W/cm$^2$ intensity and the corresponding extracted profiles of the attosecond pulses from the high-order harmonic spectrum (right panels): (a) the x component of laser field with carrier-envelope phase $\phi =0 $ (b) circularly polarized laser pulses with carrier-envelope phase $\phi =0 $ (c) circularly polarized laser pulses with carrier-envelope phase $\phi =\pi/2 $.
(d) The Morlet wavlet time profiles for H$_2^+$ molecule in a laser field, formed by combination of a left and a right-hand elliptically polarized Gaussian pulse with $\tau_p=T_d=110$ a.u. at 800 nm wavelength ($\omega=0.057$ a.u.) and $I$=8.6$\times 10^{14}$ W/cm$^2$ intensity with carrier-envelope phase $\phi =\pi/2$ and $\epsilon=0.35 $.\label {fig:q}}		
\end{figure*}
According to the above equation, introducing ellipticity to the input pulse makes the gate narrower. The calculated time profile for laser pulses with $\tau_p=T_d=220$ a.u. and $\phi=\pi/2$ is shown in Fig. 3(a). In present work, $\epsilon=0.35 $ is considered. The corresponding field amplitude is illustrated in Fig. 3(c). The amplitude of circularly fields, implemented in the previous simulations, is presented in Fig. 3(d) for comparison. As it has shown in bottom panels of Fig. 3, in this simulation the peak intensity is chosen $I$ = 8.6 $\times 10^{14}$ W$/$cm$^2$ so that the driving field strength inside the gate is the same as in the previous simulations for  H$_2^+$ molecule in superposition of circularly polarized laser pulses.
As it is apparent from the Fig. 3(c), due to the introduced ellipticity, the gating field is much stronger outside the gate, which suppresses the harmonic generation in the gate more effectively than in the case where the components of the initial pulse are circularly polarized. As it can be seen from Fig. 3(a), only the short trajectory of the main pulse is survived. The calculated attosecond pulses created by superposing the harmonics from the 60th to the 90th orders is presented in Fig. 3(b). 
As one can see only a single attosecond pulse is left with duration of 112 attosecond. Although, in contrast with the two generated pulses in the superposition of circularly polarized pulses with $\phi=\pi/2$, only a single attosecond pulse has survived here, but due to the depletion of the ground state population before the gate, the survived signal has a lower intensity than main signal in circularly polarized pulses.
This suggests that by making the gate of the circularly polarized pulses shorter, one could generate an intense isolated attosecond pulse.\\
\hspace*{3 mm}In this part, the simulations are repeated using the 110 a.u. ($\sim$ 2.5 fs) circular pulses with a 110 a.u. delay. According to the Eq. (12), for laser pulses with $\tau_p=T_d =110 $ a.u., the gate width is shorter than the spacing between two adjacent attosecond pulses. The calculated time profiles and corresponding calculated attosecond pulses are presented in Fig. 4(a)-4(d). There are fewer pulses in the gate part of the time profile in comparison with $220$ a.u. pulses, as the gate width for pulses with $\tau_p=T_d=110$ a.u. is shorter.
In comparison with $\tau_p=T_d=220$ a.u., polarization gating with a shorter gate, reduced the number of pulses in the pulse train from three to one. The duration of the survived pulses for circularly polarized laser pulses with $\phi=0$ and $\phi=\pi/2$, and elliptically polarized fields with  $\phi=\pi/2$, are calculated 155, 111 and 111 attosecond respectively as it is presented in right panels of Fig. 4(a)-4(d).
As it can be seen from the comparison of Fig . 3(b) and Fig. 4(c), the isolated attosecond pulse in the superposition of circularly polarized pulses with $\tau_p=T_d=110$ a.u. and $\phi=\pi/2$, has a higher intensity than one generated in combination of elliptically polarized pulses with 220 a.u. duration time and delay. Due to the pulse generation in the nonlinear part of the field, where the $\xi>0.2$, the generated isolated attosecond pulse in the circularly polarized pulses with $\phi=0$, Fig. 4(b), has a time-dependent ellipticity.
\section{Conclusions}
Single attosecond pulse generation by polarization gating of high-order harmonic emission from H$_2^+$ molecule, in superposition of a left and a right-hand circularly polarized Gaussian pulse, is studied and compared to the results from polarization gating of HHG in the superposition of elliptically polarized pulses.
In conclusion our quantum-mechanical calculations show that by polarization gating, one can reduce the number of attocsecond pulses in the pulse train of a linearly polarized pulse with 220 a.u. ($\sim$ 5 fs) duration time to one or two attosecond pulses.
In order to generate a single isolated attosecond pulse in superposition of polarized pulses with 220 a.u. duration, the H$_2^+$ ion should be driven by elliptically polarized laser fields.
In superposition of circularly laser fields, a single attosecond pulse is generated only when the duration time of initial pulses is chosen shorter than 220 a.u.. It is found that in 110 a.u. ($\sim$ 2.5 fs) circularly polarized pulses, a high-intensity single attosecond pulse could be obtained by selecting appropriate carrier-envelope phase.

\clearpage
\end{document}